# Diffusion controlled oriented growth of a nano porous material by the Kirkendall effect

Aloke Paul,
Department of Materials Engineering, Indian Institute of Science, Bangalore, India
E-mail: aloke@materials.iisc.ernet.in, Tel.: +918022933242, Fax: +91802360 0472

Nanotubes and nanoporous materials are being made by solid state diffusion utilizing the Kirkendall effect. Recently published manuscripts have shown that the product phase grows with an oriented microstructure. Based on the requirement of relative mobilities of the components to produce a good quality nanoporous material, it is shown that the oriented microstructure will grow invariably in most of the systems.

Formation of voids due to Kirkendall effect during diffusion controlled growth of a product phase between two dissimilar materials is a major concern in many applications such as electro-mechanical contacts in flip chip bonding [1] or multifilamentary $Nb_3Sn$ superconductor wire [2]. On the other hand, utilizing this adverse side of the Kirkendall effect, nano porous materials [3] and nano tubes [4] are being produced. Initially Aldinger [5] indicated that this phenomenon could be useful to produce porous structure by solid-state diffusion process. Much recently, numerous studies have been published producing these nano materials utilizing this concept. These can be broadly divided into two types. In one type, nano rod/wire of material A is coated with another material B such that a compound AB is produced at the interface by solid-state diffusion. In the formation of nano tubes, as shown in Figure 1a, materials are chosen such that A has much higher diffusion rate compared to the negligible diffusion rate of B through the AB product phase. Further, amounts of A and B are chosen such that both are consumed almost completely to produce a single-phase product layer. In another type, as shown in Figure 1b, nano rods/wires of material A are oxidized in oxygen such that oxide nano tubes are grown after complete consumption of A. Therefore, in both the cases, core materials diffuse out through the product phase to react with another material at the AB/B or oxide/oxygen interface for the growth of the product phase. In the latter example, recently, an oriented microstructural growth of the product phase is reported during the production of Cu-oxide layer from Cu nano wires [6], as shown in Figure 2a. This is even clearly seen in a high resolution transmission electron micrograph of a NiO layer during the production of nano porous structure from Ni nano particles [7], as hown in Figure 2b. In fact, oriented microstructure will invariably grow in both types as explained below.

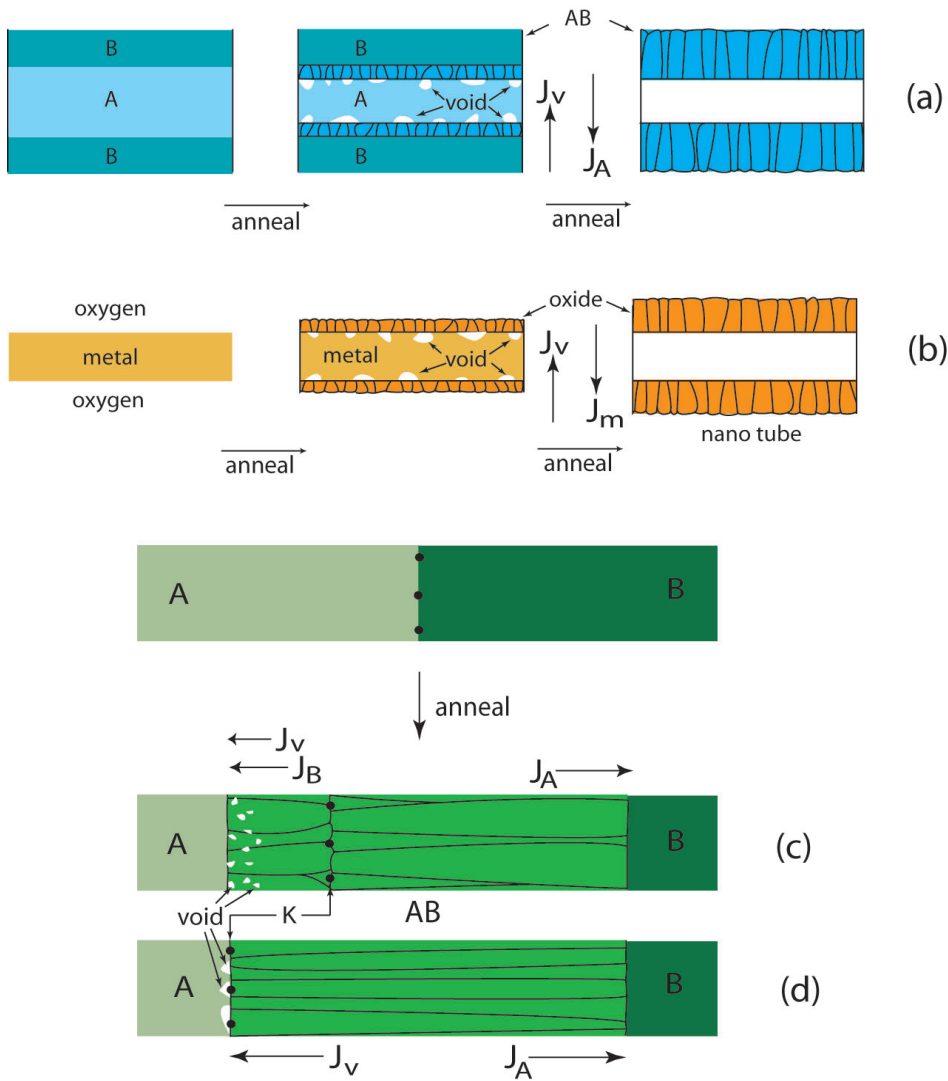

Figure 1: Formation of nano tube between (a) A-B and (b) metal-oxygen is explained. Microstructural evolution between A/B for (c) $D_A > D_B$ and $D_A \gg D_B$ are explained. "K" denotes the location of the Kirkendall marker plane.

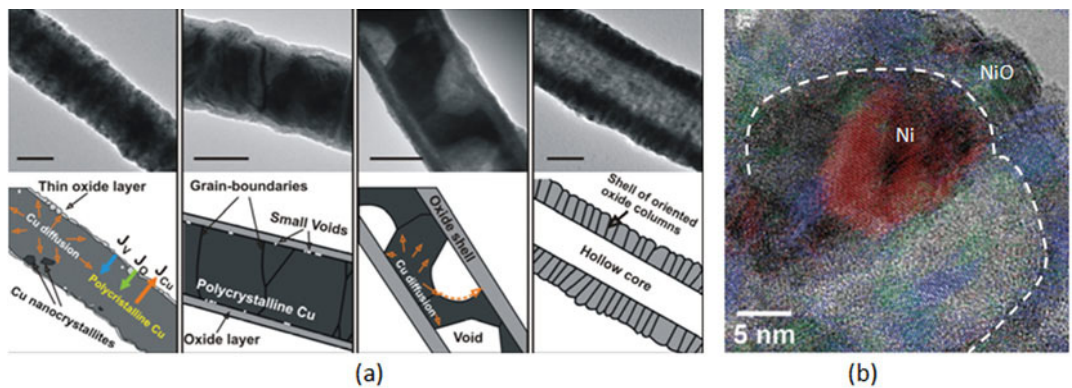

Figure 2: Growth of Cu-oxide with an oriented microstructure by the Kirkendall effect [6] (permission granted to reuse the figure) and (e) Growth of NiO with an oriented microstructure [7] (permission granted to reuse the figure). A dotted line is drawn to guide the interface between Ni and NiO.

This could be understood based on a physico-chemical approach relating the interdiffusion of components and microstuctural evolution in an interdiffusion zone [8]. The similar discussion can be adopted for the explanation of the oriented microstructure, as explained in Figure 1c. For example, we consider a diffusion couple A/B in which compound AB is grown. If both A and B diffuse with reasonable rate through the product phase, then A will diffuse from interface I to interface II and react with B to produce the product layer. Similarly, B will diffuse to interface I and react with A to produce the same product phase. Since the product phase grows with a different crystal structure than A or B, this should nucleate and grow differently from two different interfaces. Therefore, the location of the Kirkndall marker (K) plane can be easily detected by the presence of a duplex morphology [8] because of growth of two sublayers. Since growth of the phase layer is because of one dimensional flow, grains should be columnar in nature, as we have already seen in many diffusion couples [9-11]. The difference in diffusion rates of the components in a compound are dictated by many factors such atomic arrangements and differet types of defects present [8]. There could be a situation, as shown in Figure 1d, in which A has much higher diffusion rate compared to very negligible diffusion rate of B. In this case, the product phase will grow only at the interface II, since B does not diffuse to interface I for the growth from this interface. Therefore, it will have a uniform morphology and marker plane will be found at the A/AB interface.

If we consider the diffusion rates of both the components are comparable but $D_A > D_B$, as shown in Figure 1c, then the marker plane (K) will move towards A relative to the initial contact plane bewteen A and B, since more material of A (i.e. flux $J_A$) diffuse towards interface II than the diffusion of material B (i.e. flux $J_B$) towards interface I. Therefore, the resultant flux of vacancies, $J_v$ (= $J_A - J_B$) flow towards interface I [8, 12]. When these vacancies are absorbed, marker plane moves towards this interface because of shrinkage. Subsequently, there will be an expansion at the other side. On the other hand, if all the vacancies are not absorbed, voids grow by super saturation of vacancies in the region between the interface I and the marker plane [11]. These will be spread over this sublayer since phase grows along with the formation of voids at the interface I because of diffusion of B. This kind of behaviour is found in Cu-Sn system [13, 14]. In the second example in Figure 1d, where $D_A \gg D_B$ and if vacancies are not absorbed at the interface I, voids will grow at this interface. These will not be spread in the interdiffusion zone since there is no growth of the product phase from this interface because of diffusion of B.

Therefore, if the materials are chosen such that $D_A >> D_B$ and vacancies are not absorbed, voids will grow at the interface. Once the whole of A diffuse out to produce the product phase, voids coalescence to form a nano tube or nano porous material. Subsequently, the product phase will have an oriented columnar microstructure since it grows from one interface only. Moreover, the product phase should not nucleate repeatedly. In most of the systems, this is the case because of high nucleation barrier and diffusing components prefer to join with the existing grains.

To summarize, for the production of a good quality nanotube, eventually the material made of core only should be able to diffuse through the product phase. This will invariably produce a nano tube with an oriented microstructure. If the other component *i.e.* B in the A-B system or oxygen in the A/oxygen system diffuse though the product phase, this will have voids in one of the sublayers with duplex morphology in the interdiffusion zone.